\documentstyle[preprint,aps]{revtex}

\newcommand{\beq}{\begin{equation}}
\newcommand{\eeq}{\end{equation}}
\newcommand{\beqa}{\begin{eqnarray}}
\newcommand{\eeqa}{\end{eqnarray}}
\newcommand{\bef}{\begin{figure}}
\newcommand{\enf}{\end{figure}}
\newcommand{\bdis}{\begin{displaymath}}
\newcommand{\edis}{\end{displaymath}}

\newcommand{\rumore}{\mbox{\boldmath $\eta$}}
\newcommand{\der}{\partial}

\begin{document}

\title{Dispersion of passive tracers in closed basins: beyond the diffusion
coefficient}

\author{V. Artale}
\address{ENEA, CRE Casaccia, Via Anguillarese 301, Santa Maria di Galeria 
00060 Roma, Italy}

\author{G. Boffetta}
\address{Dipartimento di Fisica Generale, Universit\`a di Torino
Via Pietro Giuria 1, 10125 Torino, Italy \\
Istituto Nazionale Fisica della Materia, Unit\`a di Torino}

\author{A. Celani}
\address{Dipartimento di Ingegneria Areonautica e Spaziale,
Politecnico di Torino, \\
Corso Duca degli Abruzzi 24, 10129 Torino, Italy
\\
Istituto Nazionale Fisica della Materia, Unit\`a di Torino}

\author{M. Cencini and A. Vulpiani}
\address{Dipartimento di Fisica, Universit\`a di  Roma 
"la Sapienza", \\
Piazzale Aldo Moro 5, 00185 Roma, Italy \\
Istituto Nazionale Fisica della Materia, Unit\`a di Roma}

\date{\today}

\maketitle

\begin{abstract}
We investigate the spreading of passive tracers in closed basins.
If the characteristic length scale of the Eulerian velocities is 
not very small compared with the size of the basin the usual 
diffusion coefficient does not give any relevant information about 
the mechanism of spreading.

We introduce a finite size characteristic time $\tau(\delta)$ which 
describes the diffusive process at scale $\delta$. 
When $\delta$ is small compared with the typical length of the velocity 
field one has $\tau(\delta) \sim \lambda^{-1}$, where $\lambda$ is the maximum 
Lyapunov exponent of the Lagrangian motion.
At large $\delta$ the behavior of $\tau(\delta)$ depends on the details 
of the system, in particular the presence of boundaries, and in this limit
we have found a universal behavior for a large class of system under 
rather general hypothesis.

The method of working at fixed scale $\delta$ makes more physical
sense than the traditional way of looking at the relative diffusion
at fixed delay times.
This technique is displayed in a series of numerical experiments 
in simple flows.
\end{abstract}

\newpage
\renewcommand{\baselinestretch}{2.} 
  
\section{Introduction}
\label{sec:1}
The understanding of diffusion and transport of passive 
tracers in a given velocity field has both theoretical and practical 
relevance in many fields of science and engineering, 
e.g. mass and heat transport in geophysical flows 
(for a review see \cite{davis,davis2}),
combustion and chemical engineering \cite{Moffatt}.

One common interest is the study of the mechanisms
which lead to transport enhancement as a fluid is driven 
farther from the motionless state. 
This is related to the fact that the Lagrangian motion of 
individual tracers can be rather complex even in simple laminar flows
\cite{Ottino,lagran}.

The dispersion of passive scalars in a given velocity field is the result,
usually highly nontrivial, of two different contributions:
molecular diffusion and advection.
In particular, one can have rather fast transport, even without
molecular diffusion, in presence of {\it Lagrangian chaos \/}, 
which is the sensitivity to initial conditions of
Lagrangian trajectories.
In addition, also for a 2D stationary velocity field, 
where one cannot have Lagrangian chaos \cite{Licht}, in presence of 
a particular geometry of the streamlines the diffusion can 
be much larger than the one due only to the molecular 
contribution, as in the case of spatially periodic stationary
flows \cite{Rosenbluth,Shraiman}.

Taking into account the molecular diffusion, the motion of 
a test particle (the tracer) is described by the following Langevin
equation:
\beq
\frac{d{\bf x}}{dt}= {\bf u}({\bf x},t)+
\rumore(t),
\label{eq:langevin}
\eeq
where 
${\bf u}({\bf x},t)$ 
is the Eulerian 
incompressible velocity field at the point ${\bf x}$ 
and time $t$, $\rumore(t)$ is a Gaussian white noise 
with zero mean and
\beq
<\eta_{i}(t) \eta_{j}(t^{'}) >= 2 D_{0} \delta_{ij} \delta(t-t^{'})\,,
\label{eq:whitenoise}
\eeq
where $D_{0}$ is the (bare) molecular diffusivity.

Denoting 
$\Theta({\bf x},t)$ 
the concentration of tracers,
one has:
\beq
{\der}_{t} \Theta+ 
\left( {\bf u} \cdot \mbox{\boldmath $\nabla$} \right) \Theta=
D_{0} \,\Delta \Theta \,.
\label{eq:fokker}
\eeq

For an Eulerian velocity field  
periodic in space, or anyway  defined in infinite domains, 
the long-time, large-distance behavior of the diffusion process
is described by the effective diffusion tensor $D_{ij}^{E}$
({\it eddy-diffusivity tensor}\/):
\beq
D_{ij}^{E}= \lim_{t\rightarrow \infty} \frac {1}{2t}
<(x_{i}(t)-<x_{i}>)(x_{j}(t)-<x_{j}>)>\,,
\label{def:eddydiff}
\eeq
where now ${\bf x}(t)$ is the position of the the tracer 
at time $t$, $i,j=1,\cdots,d$ (being $d$ the spatial dimension) , and
the average is taken over the initial positions
or, equivalently, over an ensemble of test particles.
The tensor $D^{E}_{ij}$ gives the long-time, large-distance
equation for $< \! \Theta \!>$ i.e. the concentration field locally averaged
 over a volume of linear distance much larger than the
typical length $l_{u}$ of the velocity field, according to
\beq
{\der}_{t} <\Theta>= \sum_{i,j=1}^{d} D_{ij}^{E}\, 
\frac{{\der}^{2}}{\der x_{i} \der x_{j}} <\Theta>\,.
\label{eq:eddydiff}
\eeq
The above case, with finite $D_{ij}^{E}$, is
the typical situation where the diffusion, for very large 
times, is a standard diffusion process. 
However there are also cases 
showing the so-called {\it anomalous diffusion}\/: the spreading 
of the particles does not behave linearly with time but
has a power law $t^{2\nu}$ with $\nu \neq 1/2$.
Transport anomalies are, in general, indicators of the 
presence of strong 
correlation in the dynamics, even at large time and space scales
\cite{georges}.

In the case of infinite spatial domains and periodic 
Eulerian fields the powerful multiscale technique 
(also known as homogenization in mathematical literature) 
gives a useful tool for studying standard diffusion
and, with some precautions, also the anomalous situations 
\cite{BCVV}.

On the other hand we have to stress the fact that 
diffusivity tensor (\ref{def:eddydiff}) is 
mathematically well defined only in the limit of infinite times, 
therefore it gives 
a sensible result only if the characteristic length
$l_{u}$ of the velocity field is much smaller than the size 
$L$ of the domain. 

The case when $l_{u}$ and $L$ are not well separated
is rather common in many geophysical problems, e.g.
spreading of pollutants in Mediterranean or Baltic sea, 
and also in plasma physics.
Therefore it is important to introduce some other 
characterizations of the diffusion properties
which can be used also in non ideal cases.
For instance, \cite{Zambia} propose to employ exit times 
for the study of transport in basins with complicated geometry.

In Section \ref{sec:2} we introduce a characterization of
the diffusion behavior in terms of the typical time
$\tau(\delta)$ at scale $\delta$;
this allows us to define a finite size diffusion 
coefficient 
$D(\delta) \sim \delta^{2}/\tau(\delta)$.
>From the shape of $\tau(\delta)$ 
as a function of $\delta$, one can distinguish different
spreading regimes.

In Section \ref{sec:3} we present the results of numerical experiments
in closed basins and present new results
relative to the behavior of the diffusion coefficient 
near the boundary (a detailed discussion 
is in the appendix).

In Section \ref{sec:4} we summarize our results and
present conclusions and we discuss 
the possibility of treatment of experimental data 
according to the method introduced in Section \ref{sec:2}.

\section{Finite size diffusion coefficient}
\label{sec:2}
Before a general discussion let us start with a simple example.
Consider the relative diffusion of a cloud of N test particles
in a smooth, spatially periodic velocity field with characteristic
length $l_{u}$.  We assume that the Lagrangian motion is chaotic 
i.e. the maximum Lyapunov exponent $\lambda$ is positive.
Denoting with $R^{2}(t)$ the square of the typical radius of
the cloud
\beq
R^{2}(t)= 
\ll|{\bf x}_{i}(t)-\ll{\bf x}_{i}(t)\gg|^{2}\gg\,,
\label{def:disprel}
\eeq
where
\beq
\ll{\bf x}_i(t)\gg={1 \over N} \sum_{i=1}^N {\bf x}_i(t)
\eeq
we expect the following regimes to hold
\beq
\overline{R^{2}(t)} \simeq \left\{ 
\begin{array}{ll}
R^{2}(0)\exp(L(2)t) & \;\;\;\;
{\mbox {if    $\overline{R^{2}(t)}^{1/2} \ll l_{u}$}}
 \\
2 D t & \;\;\;\;
{\mbox {if    $\overline{R^{2}(t)}^{1/2} \gg l_{u}$}}
\end{array}
\label{eq:regimiperR}
\right.
\,,
\label{example1} 
\eeq
where $L(2) \geq 2\lambda$ is the generalized Lyapunov exponent
\cite{BPPV85,PV87}, $D$ is the diffusion coefficient and 
the overbar denotes the average over initial conditions.

In this paper we prefer to study the relative diffusion 
(\ref{def:disprel}) instead of the usual absolute diffusion.
For spatially infinite cases, without mean drift
there is no difference;
for closed basins the relative dispersion is,
for many aspects, more interesting than the absolute one
and, in addition, the latter is dominated by 
the sweeping induced by large scale flow.

Furthermore we underline 
that although the dynamics of the ocean circulation is dominated
by large mesoscale gyres, the smaller scales 
activities within the gyres
control important local phenomena as deep water 
formation in North Atlantic and in Mediterranean 
basin \cite{marshal}.
Therefore the study of relative diffusion could be 
relevant to describe this small-scale motion
and can give crucial informations on the way 
to parameterize the subgrid scales 
in ocean numerical global model \cite{garret}. 

Another, at first sight rather artificial, way to describe
the above behavior is by introducing the ``doubling 
time\/'' $\tau(\delta)$ at scale $\delta$ as follows:
we define a series of thresholds $\delta^{(n)}= r^{n} \delta^{(0)}$,
where $\delta^{(0)}$ is the initial size of the cloud, defined according
to (\ref{def:disprel}), and then we measure the time $T(\delta^{(0)})$ 
it takes for the growth
from $\delta^{(0)}$ to $\delta^{(1)}= r \delta^{(0)}$, and so on
for $T(\delta^{(1)})\,,\;T(\delta^{(2)})\,,\ldots$
up to the largest scale under consideration.
For the threshold rate $r$ any value can be chosen but too large ones
might not separate different scale contributions, 
though strictly speaking the term ``doubling time''
refers to the threshold rate $r=2$.

Performing ${\cal N} \gg 1$ experiments with
different initial conditions for the cloud, we define the 
typical doubling time $\tau(\delta)$ at scale 
$\delta$ as
\beq
\tau(\delta) = < T(\delta) >_e =\frac{1}{{\cal N}}
 \sum_{i=1}^{{\cal N}} T_{i}(\delta)\,.
\label{def:taudelta}
\eeq
Let us stress the fact that the average 
in (\ref{def:taudelta}) is different from the usual
time average.

From the average doubling time we can define the finite size 
Lagrangian Lyapunov exponent as
\beq
\lambda(\delta)=\frac{\ln r}{\tau(\delta)}\,,
\eeq
which is a measure of the average rate of separation of two
particles at a distance $\delta$. Let us remark that $\lambda(\delta)$
is independent of $r$, for $r \rightarrow 1^{+}$. 
For very small separations (i.e. $\delta \ll l_u$) one recovers the standard 
Lagrangian Lyapunov exponent $\lambda$,
\beq
\lambda=\lim_{\delta \rightarrow 0} \frac{1}{\tau(\delta)}
\ln r\,.
\label{def:liapfromtau}
\eeq
See \cite{ABCPV} for a detailed discussion about 
these points.
In this framework the
finite size diffusion coefficient $D(\delta)$ dimensionally turns out to be
\beq
D(\delta)=\delta^{2}\lambda(\delta)\,.
\label{def:fsd}
\eeq
Note the absence of the factor $2$, as one can expect by
the definition (\ref{def:eddydiff}), in the denominator of
$D(\delta)$ in equation (\ref{def:fsd}); this is due 
to the fact that $\tau(\delta)$ is a difference of times.
For a standard diffusion process $D(\delta)$ approaches the diffusion
coefficient $D$ (see eq. (\ref{eq:regimiperR})) in the limit of 
very large separations ($\delta \gg l_u$). This result stems from 
the scaling of the doubling times $\tau(\delta) \sim \delta^2$ for 
normal diffusion. 

Thus the finite size Lagrangian Lyapunov exponent $\lambda(\delta)$, or
its counterpart $D(\delta)$, embody the asymptotic behaviors 
\beq
\lambda(\delta) \sim \left\{ 
\begin{array}{ll}
\lambda & \;\;\;\;
{\mbox {if    $\delta \ll l_{u}$}}
 \\
D/\delta^{2} & \;\;\;\;
{\mbox {if    $\delta \gg l_{u}$}}
\end{array}
\right.
\,,
\label{eq:regimipertau} 
\eeq
One could naively conclude, matching the behaviors 
at $\delta \sim l_{u}$, that $D \sim \lambda l_{u}^{2}$.
This is not always true, since one can have a rather large range
for the crossover due to the 
fact that nontrivial correlations can be present in 
the Lagrangian dynamics \cite{FV89}.

Another case where the 
behavior of $\tau(\delta)$ as a function of $\delta$ 
is essentially well understood
is 3D fully developed turbulence. 
For sake of simplicity we neglect intermittency 
effects. 
There are then three different ranges:
\begin{enumerate}
\item 
$\delta \ll \eta ={\mbox {Kolmogorov length}}$ :
$1/\tau(\delta) \sim \lambda$;
\item
$\eta \ll \delta \ll l={\mbox{ typical size of the 
energy containing eddies}}$: 
from the Richardson law 
$\overline{R^{2}(t)} \sim t^{3}$ 
one has 
$1/\tau(\delta) \sim \delta^{-2/3}$;
\item
$\delta \gg l$ : usual diffusion behavior 
$1/\tau(\delta) \sim \delta^{-2}\,.$
\end{enumerate}

One might wonder that the proposal to introduce
the time $\tau(\delta)$ is just another way 
to look at
$\overline{R^{2}(t)}$ as a function of $t$.
This is true only in limiting cases, when
the different characteristic lengths are 
well separated and intermittency is weak.
In \cite{previouswork1,previouswork2,sabot} rather 
close techniques are used for the computation of
the diffusion coefficient in nontrivial cases.

The method of working at fixed scale $\delta$,
allows us to extract the physical information at that spatial
scale avoiding unpleasant troubles of the method of
working at a fixed delay time $t$.
For instance, if one has a strong intermittency, and this is a rather
usual situation, $R^{2}(t)$ as a function of 
$t$ can appear very different in each realization.
Typically one can have, see figure \ref{fig1}a,
different exponential rates of growth for different
realizations, producing a rather odd behavior
of the average $\overline{R^{2}(t)}$ without
any physical meaning. For instance in figure \ref{fig1}b we show
the average  $\overline{R^{2}(t)}$ versus time $t$; at large times we
recover the diffusive behavior but at intermediate times 
appears an apparent anomalous regime which is only due to 
the superposition of exponential and diffusive contributions
by different samples at the same time.
On the other hand exploiting the tool of doubling times one has 
an unambiguous result (see figure \ref{fig1}c).

Of course the interesting situations are those where
the different characteristic lengths ($\eta\,,\;l\,,\;L$) 
are not very different and therefore each 
scaling regime for $\overline{R^2(t)}$ is not well evident.

\section{Numerical results}
\label{sec:3}
Here we present some numerical experiments 
in simple models with
Lagrangian chaos in the zero molecular diffusion limit.
Before showing the results, we describe the numerical 
method adopted.

We choose a passive tracers trajectory having a chaotic behavior, 
i.e. with a positive maximum Lyapunov 
exponent, computed by using standard algorithms \cite{BeneGalg}.
Then we place $N-1$ passive tracers around the first one
in a cloud of initial size 
\beqa
R(0)=\delta(0)=\delta^{(0)}\,,
\nonumber
\eeqa 
with $R(0)$ defined by equation (\ref{def:disprel}).  
In order to have average properties we repeat this procedure 
reconstructing the passive cloud around the last
position reached by the reference chaotic tracer in the previous 
expansion.
This ensures that the initial expansion of the cloud 
is exponential
in time, with typical exponential rate equal to the
Lyapunov exponent.

Further we define a series of thresholds $\delta^{(n)}=r^{n}\delta^{(0)}$
(as described in Section 2) 
$n=1,\cdots,n_{max}$ and we measure the time $T_{n}$ 
spent in  expanding from $\delta^{(n)}$ to $\delta^{(n+1)}\,$.
The value of $n_{max}$ has to be chosen in such a way that
$\delta^{(n_{max})}\sim \delta_{max}$, where $\delta_{max}$
corresponds to the uniform distribution of the tracers in the basin
 (see forthcoming discussion and the Appendix). Each realization stops
when $\delta(t)=\delta^{(n_{max})}$.

Therefore following \cite{ABCPV} we define a scale dependent
Lagrangian Lyapunov exponent as:
\beq
\lambda(\delta^{(n)}) = \frac{1}{<{T_{n}}>_e} \ln r =
\frac{1}{\tau(\delta^{(n)})} \, \ln r.
\label{def:lambdadidelta}
\eeq
In equation (\ref{def:lambdadidelta}) we have implicitly assumed that
the evolution of the size $\delta(t)$ of the cloud is continuous in time.
This is not true in the case of discontinuous processes such as maps or
in the analysis of experimental data taken at 
fixed delay times.
Denoting $T_{n}$ the time to reach size 
$\tilde{\delta} \geq \delta^{(n+1)}$ from $\delta^{(n)}$ ,
now $\tilde{\delta}$ is a fluctuating quantity,
 equation (\ref{def:lambdadidelta}) has to be modified as follows 
\cite{ABCPV}:
\beq
\lambda(\delta^{(n)}) = \frac{1}{<{T_{n}}>_e}   
\left< \ln \left( \frac{\tilde{\delta}}{\delta^{(n)}}\right) \right>_e
\,.
\label{def:lambdadidelta1}
\eeq

In our numerical experiments we have the regimes 
described in sect. 2: exponential regime
, i.e. $\lambda(\delta)=\lambda$, and diffusion-like regime
i.e. $\lambda(\delta)=D/\delta^{2}$, at least if the size $L$ 
of the basin is large enough.

For cloud sizes close to the saturation value $\delta_{max}$
we expect the following behavior to hold for a broad class 
of systems:
\beq
 \lambda(\delta)=\frac{D(\delta)}{\delta^{2}} \propto
\frac{(\delta_{max}-\delta)}{\delta} \,.
\label{eq:nearbound}
\eeq
The constant of proportionality
is given by  the second eigenvalue of the 
Perron-Frobenius operator which is related to the typical time 
of exponential relaxation of tracers' density to the uniform distribution
Actually, the analytical evaluation of this eigenvalue can be 
performed only for extremely simple dynamical systems
(for instance random walkers, as shown in the Appendix).
As a consequence the range of validity for (\ref{eq:nearbound})
can be assessed only by numerical simulation.

\subsection{A model for transport in Rayleigh-B\'enard convection}
The advection in two dimensional incompressible flows is described,
in absence of molecular diffusion, by Hamiltonian equation of motion  
where the Hamilton function is the stream function $\psi$:
\beq
\frac{dx}{dt}=\frac{\der \psi}{\der y}\,, \;\;\;
\frac{dy}{dt}=-\frac{\der \psi}{\der x}\,.
\label{eq:hamilton}
\eeq
If $\psi$ is time-dependent the system (\ref{eq:hamilton})
is non-autonomous and in general non-integrable, then
chaotic trajectories may exist. 

One example is the model introduced in \cite{gollub}
to describe the chaotic advection
in the time-periodic Rayleigh-B\'enard convection.
It is defined by the stream function:
\beq
\psi(x,y,t)=\frac{A}{k} 
\sin\left\{ k \left[ x+B \sin(\omega t)\right]\right\}
W(y)\,,
\label{eq:gollubinf}
\eeq
where $W(y)$ is a function that satisfies rigid 
boundary conditions on the surfaces $y=0$ and $y=a$ 
(we use $W(y)=\sin(\pi y/a)$).
The direction $y$ is identified with the vertical direction
and the two surfaces $y=a$ and $y=0$ are the top and bottom
surfaces of the convection cell.
The time dependent term $B\sin(\omega t)$ represents 
lateral oscillations of the roll pattern  
which mimic the even oscillatory instability \cite{gollub}.

Trajectories starting near the roll
separatrices could have positive Lyapunov exponent and thus
display chaotic motion and diffusion in the x direction. 
It is remarkable that in spite of the simplicity of the model,
the agreement of the numerical results with experimental ones is quite
good \cite{gollub}.

Defining a passive cloud in the $x$ direction (i.e. a
segment) and performing the expansion experiment described 
in the previous section
we have that, until $\delta$ is below a fraction of the
dimension of the cell, $\lambda(\delta)=\lambda$ (figure \ref{fig2}a).
For larger values of $\delta$ we have 
the standard diffusion $\lambda(\delta)=D/\delta^{2}$ 
with good quantitative agreement with the value of the 
diffusion coefficient evaluated by the standard technique, i.e.
using $\overline{R^{2}(t)}$ as a function of time $t$
(compare figure \ref{fig2}a with figure \ref{fig2}b).

To confine the motion of tracers in a closed domain,
i.e. $x \in [-L,L]$, we must slightly modify the streamfunction
(\ref{eq:gollubinf}). 
We have modulated the oscillating term
in such a way that for $|x|=L$ the amplitude of the oscillation
is zero, i.e. $B \rightarrow B \sin(\pi x/L)$ with $L=2\,\pi n/k$
($n$ is the number of convective cells).
In this way
the motion is confined in $[-L,L]$.

In figure \ref{fig3} we show $\lambda(\delta)$ for two values of $L$. 
If $L$ is large enough one can well see the three regimes, 
the exponential one, the diffusive one and the saturation given
by equation (\ref{eq:nearbound}).
Decreasing $L$ the range
of the diffusive regime decreases, and for small values of
$L$ it disappears. 

\subsection{Modified Standard Map}
One of the simplest deterministic dynamical system displaying both
exponential growth of separation for close trajectories 
and asymptotic diffusive behavior 
is the standard (Chirikov - Taylor) mapping \cite{Chi79}.
It is customarily defined as 
\beq
\left\{
\begin{array}{ll}
x_{n+1}=x_n+K \sin y_n & \\
 y_{n+1}=y_n+x_{n+1}& \mbox{ mod $2 \pi $}
\end{array}
\right.
\label{eq:standard}
\eeq
This mapping conserves the area in the phase space.
It is widely known that for large enough values of the 
nonlinearity strength parameter $K \gg K_c \simeq 1$ the motion
is strongly chaotic in almost all the phase space.
In this case the standard map, in the $x$-direction
 mimics the behavior of a one-dimensional random walker,
still being deterministic, and so one expects the behavior of
$\lambda(\delta)$ to be quite similar to the one already 
encountered in the model for Rayleigh-B\'enard convection
without boundaries. 
Numerical iteration of (\ref{eq:standard}) for a cloud of particles
clearly shows the two regimes described in 
(\ref{eq:regimipertau}), similar to that showed for the 
model discussed in the previous section.

We turn now to the more interesting case in which the domain is limited
by boundaries reflecting back the particle. 
To achieve the confinement of the trajectory
inside a bounded region we modify the standard map in the following
way
\beq
\left\{
\begin{array}{ll}
x_{n+1}=x_n+K f(x_{n+1})\sin y_n & \\ 
y_{n+1}=y_n+x_{n+1}-K f'(x_{n+1}) \cos y_n & \mbox{ mod $ 2 \pi$}.
\end{array}
\right.
\label{eq:modified}
\eeq
where $f(x)$ is a function which has its only zeros in $\pm L$.  
Since the mapping is defined in implicit form, 
the shape of $f$ must be chosen in such a way to assure
a unique definition for $(x_{n+1},y_{n+1})$ given $(x_n,y_n)$. 
For any $f$ fulfilling this request the mapping 
(\ref{eq:modified}) conserves the area.
A trial choice could be
\beq
f(x)=
\left\{
\begin{array}{ll}
1 & |x|<\ell \\
\begin{displaystyle}
{L-|x| \over L-\ell}
\end{displaystyle}
 & \ell<|x|<L
\end{array}
\right.
\label{eq:f}
\eeq
Strictly speaking this is not quite an appropriate choice, since it renders
the map discontinuous at $|x| = \ell$, but 
this is an irrelevant point and it is easy to bypass this
obstacle by assuming a suitably smoothed version of (\ref{eq:f}).

Performing the doubling times computation (\ref{def:taudelta})
one recovers both
the exponential and diffusive regimes for 
$\lambda(\delta)$, and in addition one has the saturation regime
(\ref{eq:nearbound}). Figure \ref{fig4} shows the behavior of
the scale dependent diffusion coefficient $D(\delta)$ (\ref{def:fsd}).
Approaching the saturation value $\delta_{max}$ the diffusion 
coefficient quickly drops to zero, following the asymptotic law 
(\ref{eq:nearbound}) derived in the appendix.
The qualitative behaviors in figure \ref{fig4} do not depend on the details 
of the function $f$. 
                   
\subsection{Point vortices in a Disk}
As another example, we consider the  two-dimensional  
time-dependent flow generated by the motion of $N$ point vortices 
in a closed domain \cite{Aref}.
For a unitary disk the positions of the vortices $(x_i=r_i \cos \,\theta_i
,y_i=r_i \sin \,\theta_i)$, with circulation
$\Gamma_i$, evolve according to the Hamiltonian dynamics 
\begin{equation}
\dot{x_i} = {1 \over \Gamma_i} {\partial H \over \partial y_i}, \: \: \:
\dot{y_i} = - {1 \over \Gamma_i} {\partial H \over \partial x_i}
\end{equation}
where the Hamiltonian is
\beq
H = - {1 \over 4\pi} \sum_{i>j} \Gamma_i \Gamma_j \log \left[
{r_i^2+r_j^2-2 r_i r_j \cos (\theta_i-\theta_j) \over 1 + r_i^2 r_j^2 -
2 r_i r_j \cos (\theta_i-\theta_j)} \right] +
{1 \over 4 \pi} \sum_{i=1}^N \Gamma_i^2 \log (1-r_i^2)
\label{}
\end{equation}

Passive tracers evolve according to (\ref{eq:hamilton}) with $\psi$ given
by
\begin{equation}
\psi(x,y) = - {1 \over 4\pi} \sum_{i}^{N} \Gamma_i
\log \left[{r^2+r_i^2-2 r r_i \cos(\theta-\theta_i) \over
1 + r^2 r_i^2 - 2 r r_i \cos(\theta-\theta_i)} \right]
\label{}
\end{equation}
where $(x=r \cos \,\theta,y=r \sin \,\theta)$ denote the tracer
position.

Figure \ref{fig5} shows the relative diffusion as a function of 
time in a system with 4 vortices. 
Apparently there is an intermediate regime of anomalous diffusion.
On the other hand from figure \ref{fig6} one can see rather clearly 
that, with the method of working at fixed scale, only
two regimes survive: the exponential one and that one 
due to the saturation. 
Comparing figure \ref{fig5} and figure \ref{fig6} one understands that
the mechanism described in Section 2 
has to be held for responsible of this spurious anomalous diffusion.
We stress the fact that these misleading behaviors are 
due to the superposition of different regimes and that 
the method of working at fixed scale has the advantage 
to eliminate this trouble.

The absence of the diffusive range 
$\lambda(\delta) \sim \delta^{-2}$
is due to the fact that the characteristic 
length of the velocity field, which is comparable with 
the typical distance between two close vortices, is not 
much smaller than the size of the basin.

\section{Conclusions}
\label{sec:4}
In this paper we investigated the relative dispersion of passive tracers
in closed basins. Instead of the customary approach based on 
the average size of the cloud of tracers as a function of time,
we introduced a typical inverse time $\lambda(\delta)$ which 
characterizes the diffusive process at fixed scale $\delta$.

For very small values of $\delta$, $\lambda(\delta)$ coincides with the
maximum Lagrangian Lyapunov exponent which is positive in
the case of chaotic Lagrangian motion.
For larger $\delta$ the shape of $\lambda(\delta)$ 
depends on the detailed mechanism of spreading which is given
by the structure of the advecting flow, which is in turn conditioned 
by the presence of boundaries. In the case of diffusive regime, one
expects the scaling $\lambda(\delta) \simeq \delta^{-2}$, which leads to a
natural generalization of the diffusion coefficient as
$D(\delta)=\lambda(\delta) \delta^2$. 

The effectiveness of finite size quantities $\lambda(\delta)$ 
or $D(\delta)$ in characterizing the dispersion properties of
a cloud of particles is demonstrated by several numerical examples.

Furthermore, when $\delta$ gets close to its saturation value
(i.e. the characteristic size of the basin), a simple argument gives 
the shape of $\lambda(\delta)$ which is expected to be universal
with respect to a wide class of dynamical systems.

In the limiting case when the characteristic length of
the Eulerian velocity $l_u$ and the size of the basin $L$ are
well separated, the customary approach and the proposed method
give the same information. 
In presence of strongly intermittent Lagrangian motion, or when
$l_u/L$ is not much smaller than one, the traditional method
can give misleading results, for instance apparent anomalous
scaling over a rather wide time interval, as demonstrated by
a simple example.

We want to stress out that our method is very 
powerful in separating the different scales acting on diffusion 
and consequently it could give improvement about the parameterization 
of small-scale motions of complex flows.
The proposed method could be also relevant in the analysis of
drifter experimental data or in numerical models for Lagrangian
transport, in particular for addressing the question about the
existence of low dimensional chaotic flows.

\section{Acknowledgments}
We thank E. Aurell and A. Crisanti for useful suggestions and 
first reading of the paper. G.B. and A.C. thank the Istituto di
Cosmogeofisica del CNR, Torino, for hospitality.
This work was partially supported by INFN {\it Iniziativa specifica
Meccanica Statistica FI11}, by CNR (Progetto speciale coordinato
{\it Variabilit\`a e Predicibilit\`a del Clima}) and by EC-Mast contract
MAS3-CT95-0043 (CLIVAMP).

\section*{Appendix}
\appendix

In this appendix we present the derivation of the 
asymptotic behavior (\ref{eq:nearbound}) of $\lambda(\delta)$
for $\delta$ near to the saturation,
for a one dimensional Brownian
motion in the domain  $[-L,L]$, 
with reflecting boundary conditions.
The evolution of the probability density $p$ is ruled by the 
Fokker-Planck equation
\begin{equation}
{\partial p \over \partial t} = {1 \over 2} D
 {\partial^2 p \over \partial x^2}
\label{FP}
\end{equation}
with the Neumann boundary conditions
\begin{equation}
{\partial p \over \partial x}(\pm L)=0 .
\label{bc}
\end{equation}
The general solution of (\ref{FP}) is 
\begin{equation}
p(x,t)= {\displaystyle \sum_{k=-\infty}^{\infty} \hat{p}(k,0) e^{i k x}
e^{-t/ \tau_k} + c.c }\\  
\label{sol}
\end{equation}
where 
\begin{equation}
{\displaystyle\tau_k=\left( {D \over 2} {\pi^2 \over L^2} k^2  \right)^{-1}} ,
\, \, \, k=0, \pm 1, \pm 2, ...
\end{equation}
At large times $p$ approaches the uniform solution 
$p_0=1/2L$.
Writing $p$ as  $p(x,t)= p_0 + \delta p(x,t)$
we have, for $t \gg \tau_1$ ,
\begin{equation}
\delta p \sim \exp(-t/\tau_1).
\label{asi}
\end{equation} 
The asymptotic behavior for the relative dispersion
$R^2(t)$ is
\begin{equation}
R^2(t)={1 \over 2} \int (x-x')^2 p(x,t) p(x',t) \;dx\; dx' 
\end{equation}
For $t \gg \tau_1$ using (\ref{asi}) we obtain
\begin{equation}
R^2(t) \sim \left( {L^2 \over 3}-A e^{-t/ \tau_1} \right) .
\end{equation}
Therefore for  $\delta(t)=R(t)$ one has
\begin{equation}
\delta(t) \sim 
\left( {L \over \sqrt{3}} -{\sqrt{3} A \over 2 L}  e^{-t/ \tau1} \right)
\end{equation}
The saturation value of $\delta$ is 
$\delta_{max}=L/\sqrt{3}$, so 
for $t \gg \tau_1$, or equivalently for 
$(\delta_{max}-\delta)/\delta \ll 1$, we expect
\begin{equation}
{d \over dt} \ln \delta =\lambda({\delta})=
{1 \over \tau_1}{\delta_{max}-\delta \over \delta}
\label{satu}
\end{equation}
which is (\ref{eq:nearbound}).

Let us remark that in the previous argument for 
$\lambda(\delta)$ for
$\delta \simeq \delta_{max}$ the crucial point is the 
exponential relaxation to the asymptotic  uniform distribution.
In a generic deterministic chaotic system it is not
possible to prove this property in a rigorous way.
Nevertheless one can expect that this request is fulfilled 
at least in  non-pathological cases. In the terminology of 
chaotic systems the exponential relaxation to asymptotic 
distribution corresponds to have the second eigenvalue $\alpha$ of the 
Perron-Frobenius operator inside the unitary circle; now
the relaxation time is $\tau_1=-\ln|\alpha|$ \cite{Beck}.
 



\begin{figure}
\caption{
a) Three realizations of $R^2(t)$ as a function of $t$ built 
as follows:
$R^2(t)=\delta_0^2 \exp(2 \gamma t)$ if $R^2(t)<1$ and 
$R^2(t)=2 D(t-t_{\ast})$ with $\gamma=0.08, 0.05, 0.3$ and 
$\delta_0=10^{-7}, \,\; D=1.5$.
b) $\overline{R^2(t)}$ as function of $t$ averaged on 
the three realizations shown in figure 1a. The apparent anomalous
regime and the diffusive one are shown. 
c) $\lambda(\delta)$ vs $\delta$, with Lyapunov and 
diffusive regimes.
}
\label{fig1}
\end{figure}

\begin{figure}
\caption{
Lagrangian motion given by the Rayleigh-B\'enard convection model
with:
$A=0.2, \,\; B=0.4, \,\; \omega=0.4, \,\; k=1.0, \,\; a=\pi$, the number of 
realizations is ${\cal N}=2000$ and the series of thresholds 
is $\delta_n=\delta_0 r^n $ with $\delta_0=10^{-4}$ and $r=1.05$. 
a)$\lambda(\delta)$ vs $\delta$, the horizontal line indicates the Lyapunov  
exponent $\lambda=0.022$, the dashed line is $D \delta^{-2}$ with $D=0.26$.
b)$\overline{R^{2}(t)}$ as function of $t$, the line is $2D\,t$
with $D=0.26$.
}
\label{fig2}
\end{figure}

\begin{figure}
\caption{
$\lambda(\delta)$ vs $\delta$ for the same model and parameters of figure 2, 
but in a closed domain with $6$ (crosses) and $12$ (diamonds) 
convective cells. 
The lines are respectively:
(a) Lyapunov regime with $\lambda=0.017$; (b) diffusive regime 
with $D=0.021$; (c) saturation regime with $\delta_{max}=19.7$; 
(d) saturation regime with $\delta_{max}=5.7$.
}
\label{fig3}
\end{figure}

\begin{figure}
\caption{
$D(\delta)$ vs $\delta$ for the modified standard map
with $K=8$, $L=1000$ and $l=990$. 
The series of thresholds is $\delta_n=\delta_0 r^n $ with $\delta_0=10^{-4}$
and $r=2^{1/16}$.   
The horizontal line indicates 
the diffusion coefficient in the limit of infinite system, 
the dashed curve represents the saturation regime.
}
\label{fig4}
\end{figure}

\begin{figure}
\caption{
$\overline{R^{2}(t)}$ for the four vortex system
with $\Gamma_{1}=\Gamma_{2}=-\Gamma_{3}=-\Gamma_{4}=1$. The 
threshold parameter is $r=1.03$ and $\delta_{0}=10^{-4}$, 
the dashed line is the power law
$\overline{R^{2}(t)}\sim t^{1.8}$. The number of 
realizations is ${\cal N} = 2000$.
}
\label{fig5}
\end{figure}

\begin{figure}
\caption{
$\lambda(\delta)$ vs $\delta$ for the same model 
and parameters of figure 5. The horizontal line indicates 
the Lyapunov exponent ($\lambda=0.14$), 
the dashed curve is the saturation regime with $\delta_{max}=0.76$. 
}
\label{fig6}
\end{figure}

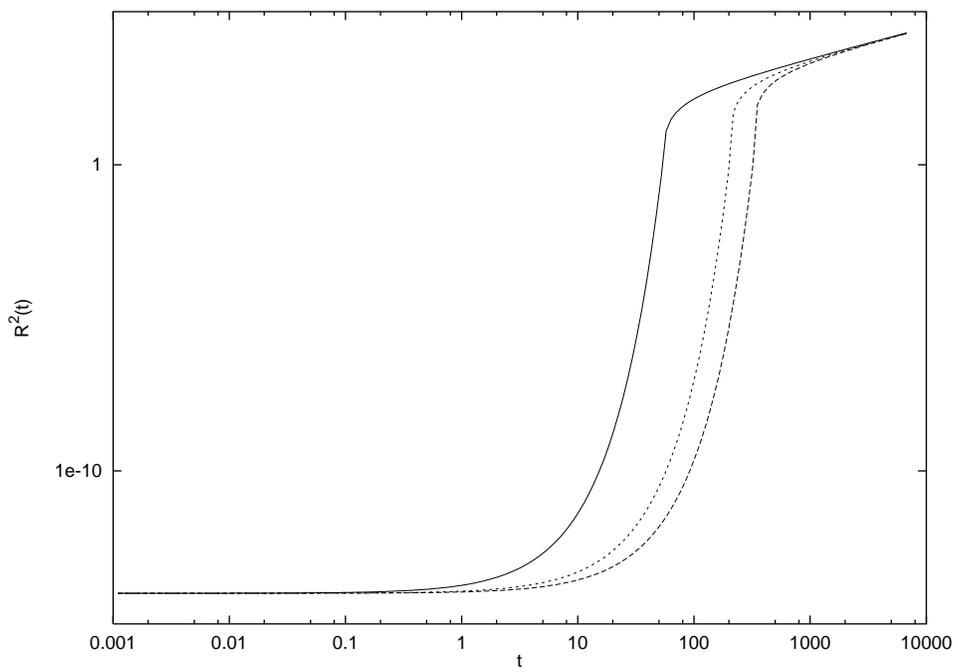
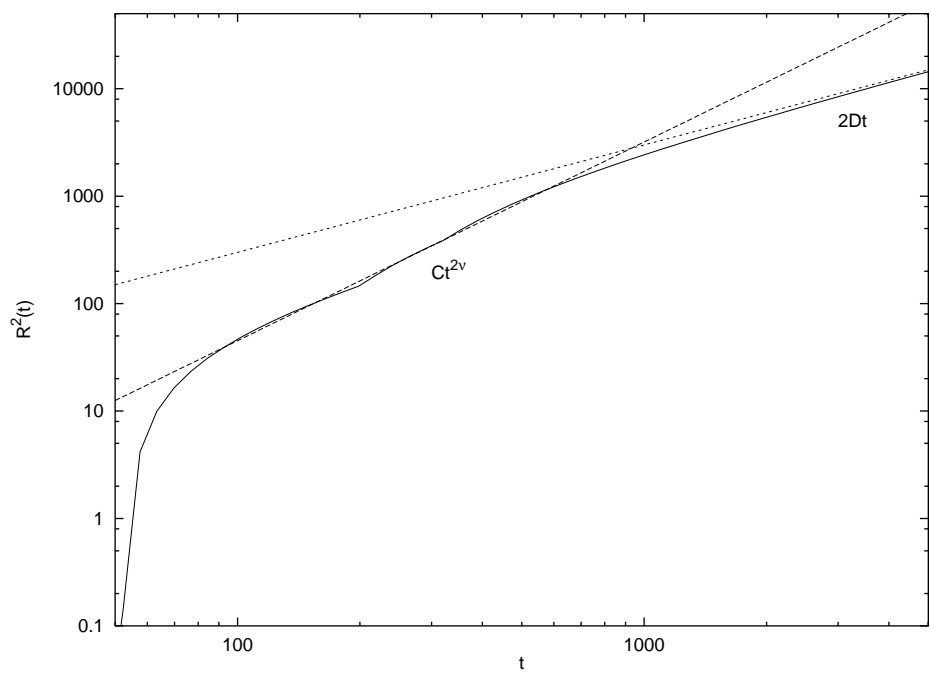
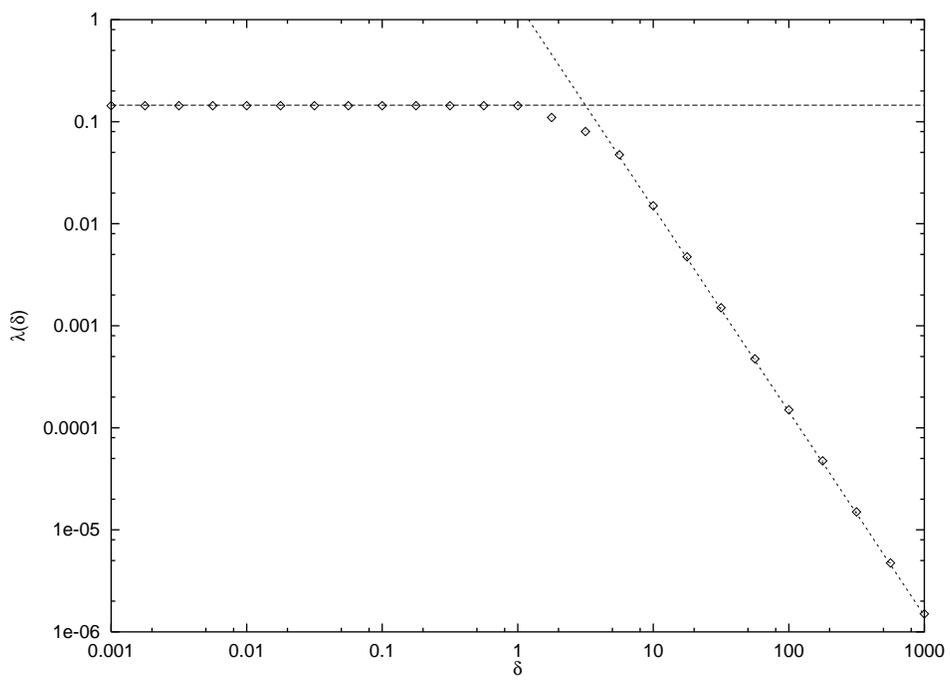
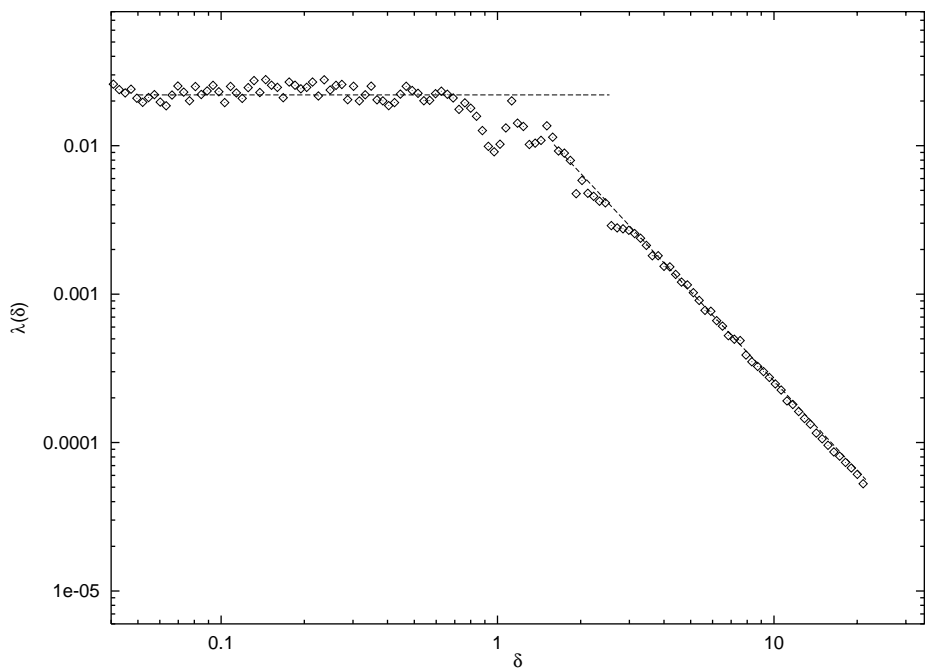
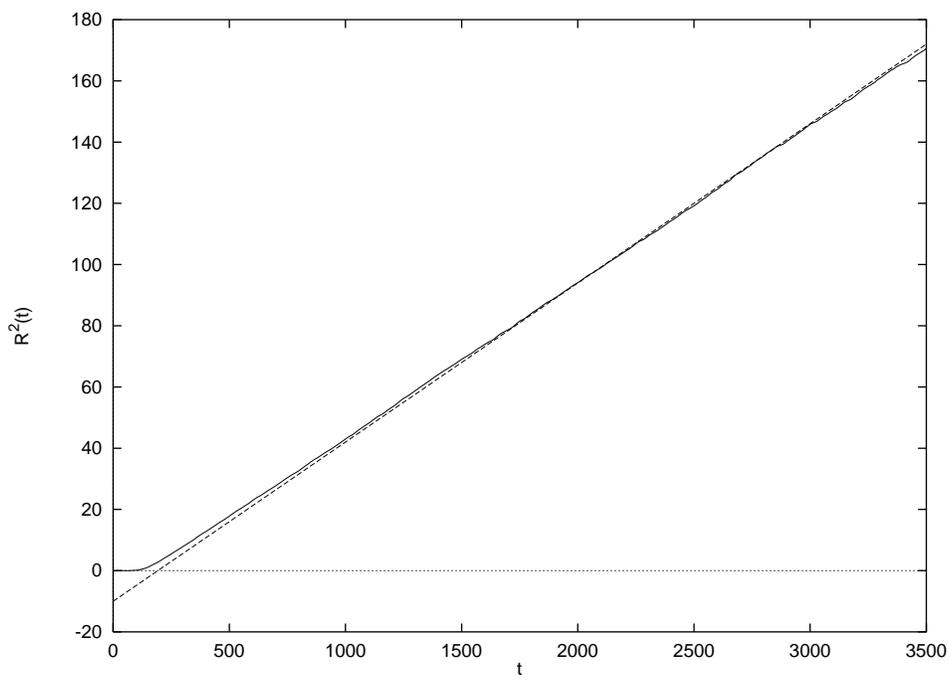
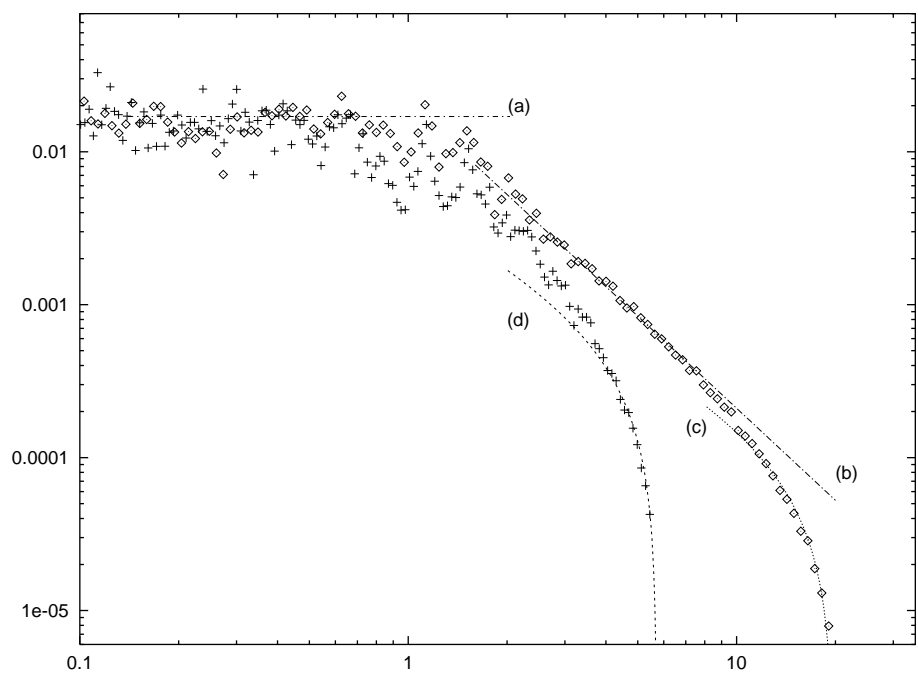
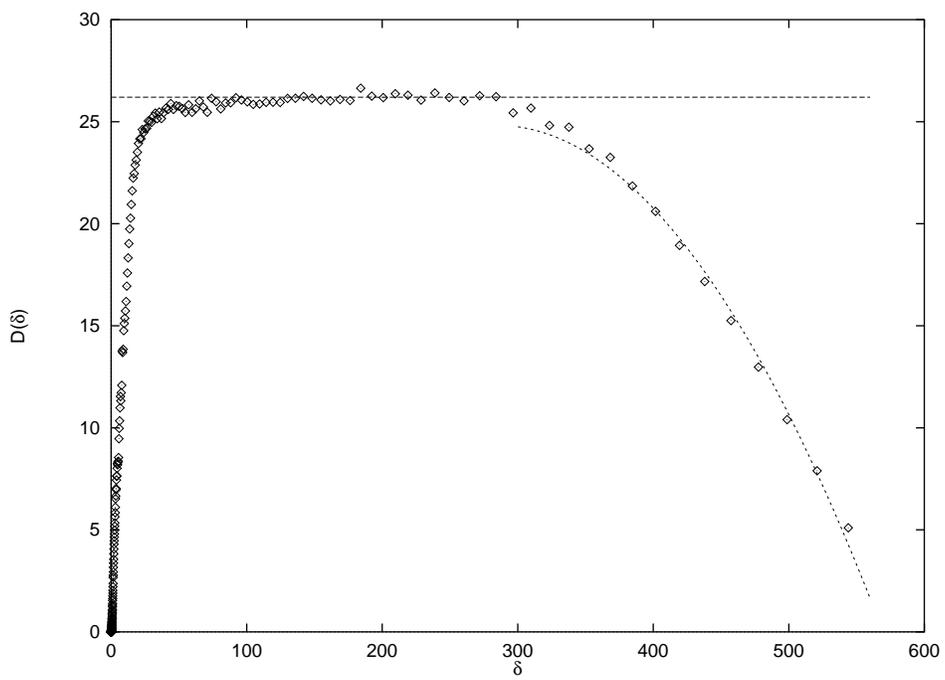
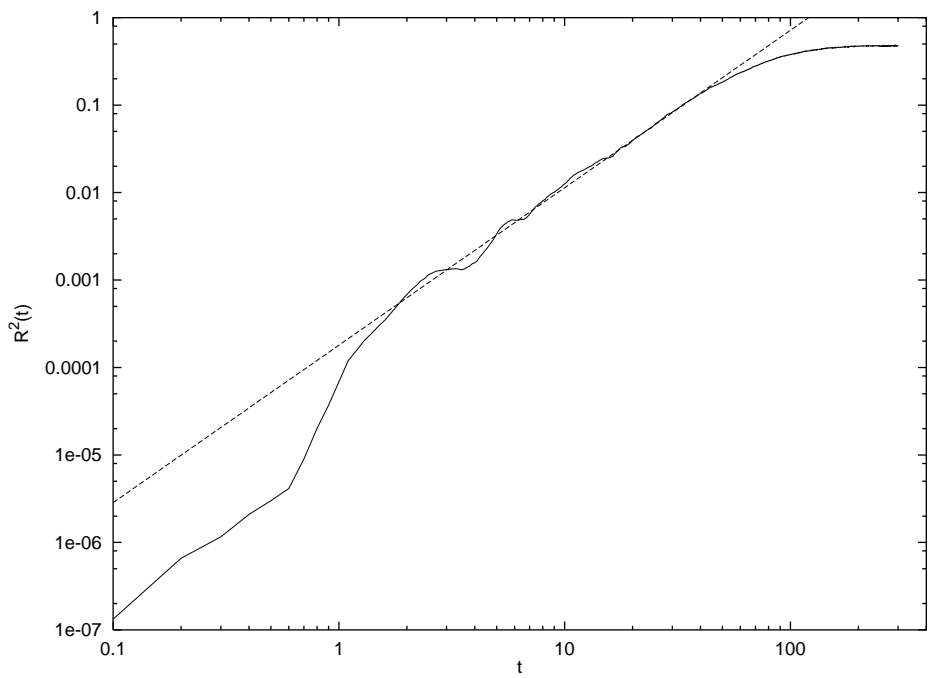
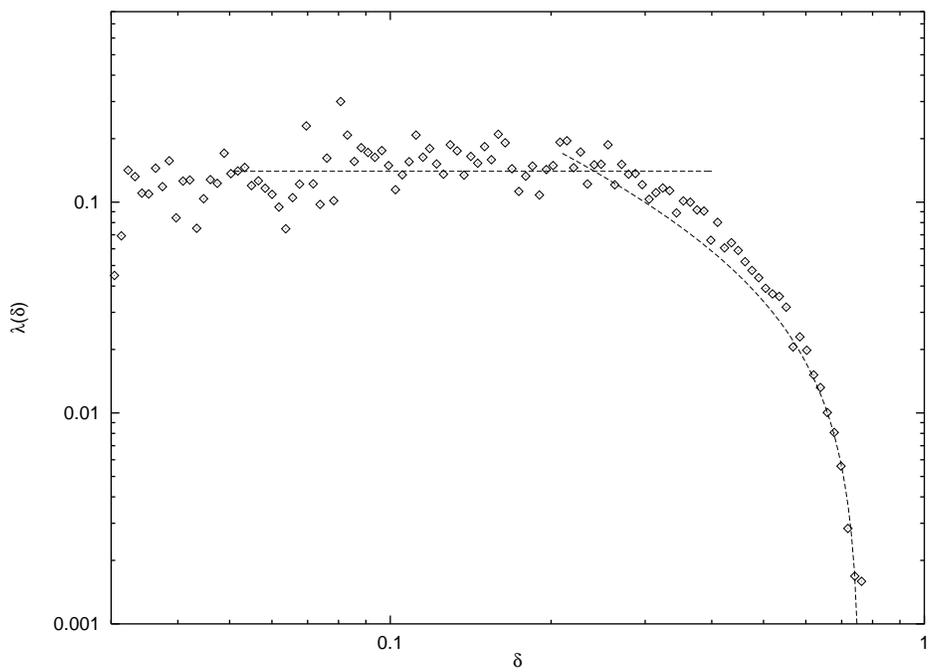
\end{document}